\documentclass[11pt]{article}

\usepackage[margin=1in]{geometry}
\usepackage{latexsym}
\usepackage{graphicx}
\usepackage{mathptmx}
\usepackage[T1]{fontenc}
\usepackage{amsmath}
\usepackage{amsfonts}
\usepackage{amssymb}
\usepackage{amsbsy}
\usepackage{amsthm}
\usepackage{placeins}
\usepackage{subcaption}
\usepackage{dsfont}
\usepackage{natbib}
\usepackage[pdftex,colorlinks=true,urlcolor=blue,citecolor=black,anchorcolor=black,linkcolor=black]{hyperref}

\newcommand{\email}[1]{\href{mailto:#1}{#1}}

\newtheoremstyle{wsc}
{3pt}
{3pt}
{}
{}
{\bfseries}
{}
{.5em}
{}

\theoremstyle{wsc}

\begin{document}

%***************************************************************************
% AUTHOR: AUTHOR NAMES GO HERE
% FORMAT AUTHORS NAMES Like: Author1, Author2 and Author3 (last names)
%
%		You need to change the author listing below!
%               Please list ALL authors using last name only, separate by a comma except
%               for the last author, separate with "and"
%

% AUTHOR: Enter the title, all letters in upper case
\title{Subtrace-Conditional Validation of Simulation Models \\ and Digital Twins}
%\title{Validating Simulation Models and Digital Twins by Conditioning on Subtraces}

% AUTHOR: Enter the authors of the article, see end of the example document for further examples
\author{
Mohammadmahdi Ghasemloo\textsuperscript{1} \and
David J. Eckman\textsuperscript{1} \and
Yaxian Li\textsuperscript{2}\\[6pt]
\small \textsuperscript{1}Dept. of Industrial and Systems Engineering, Texas A\&M University, College Station, TX, USA\\
\small \textsuperscript{2}Intuit AI, Mountain View, CA, USA
}
\date{}

\maketitle

\vspace{-12pt}

\section*{ABSTRACT}

Validating simulation models against historical output data is essential for their successful deployment in digital-twin environments. We propose a statistical validation framework in which a simulation model is repeatedly initialized from observed system states, and conditional output distributions are obtained by fixing the random primitives from a subset of stochastic input models to their observed realizations while simulating the remaining primitives. These conditional output distributions are then used in goodness-of-fit tests to validate the simulation model with respect to combinations of input models. We also develop diagnostic tools to identify the input models that most contribute to any observed misalignment between a simulation model's outputs and reality. Numerical experiments on an M/M/1 queueing system and a digital-twin-enabled simulation of a tandem queueing system demonstrate that the proposed framework can detect misspecifications in input models that may be missed by existing approaches that validate only the marginal output distribution. 

\section{INTRODUCTION}
\label{sec:intro}

Simulation models are used to support decision making in complex stochastic systems. 
The value of a simulation model to a decision maker depends on its ability to faithfully reproduce the behavior of the real system. This motivates the need for \emph{validation} procedures \cite{law2024simulation}. Some efforts take a qualitative viewpoint toward validation \cite{hua2022validation}, while others adopt a statistical perspective and provide mathematical frameworks for validation \cite{kleijnen1995verification}. A standard approach to statistically validate a simulation model is to run multiple independent replications using the estimated input models and compare the outputs with the observed real-world outputs to ensure that the input models have been estimated correctly \cite{sargent2010new,sargent2010verification}.

An application of simulation models that has gained more attention in recent years is \textit{digital twins}  \cite{pylianidis2022simulation}. A digital twin (or, more accurately, a digital shadow) is a virtual representation of a physical system that dynamically interacts with a physical system to mirror the evolving state and operational dynamics of the system \cite{huang2021survey,taylor2023enhancing}.
In a digital-twin setting, the state of the real system is observed periodically, at which time an action can be taken to alter the system.
As a result, simulation models employed as digital twins can be repeatedly initialized from observed system states.

A common challenge in validating digital twins is that the observed historical trajectories feature system outputs whose distributions are time, state, and (possibly) action dependent.  \cite{rhodes2023tracking}
establish a validation framework tailored to such environments that addresses this challenge by standardizing outputs from different periods via a probability integral transformation (PIT) approximated via simulation.
Goodness-of-fit tests are then applied to test whether the model is valid or exhibits systematic bias. \cite{he2024digital} extend this approach to the setting with multi-variate outputs by utilizing multiple hypothesis tests that assess alignment in both the marginal and joint distributions of the outputs.

The aforementioned works provide a principled and statistically grounded framework for testing the alignment between a digital twin and its physical counterpart. Nevertheless, they have limitations. 
These approaches consider all input models collectively and can, in some cases, fail to detect lower-level misalignments due to the cancellation of errors from multiple misspecified input models. These methods also do not provide guidance to the modeler about \textit{which} input models are responsible for any detected misalignment. To this end, we propose a more expansive statistical validation framework that seeks to address two fundamental questions: \textit{Does there exist any subset of input models for which the simulation model fails to reproduce the observed behavior? And what is each input model's contribution to such misalignment?}

The key innovation in our framework is the careful reuse of historical observations from the real-world input processes, which we refer to as \textit{trace data}. 
In many practical settings, one constructs digital twins of existing systems; hence, data describing the historical behavior of the physical system are often available and can be mined to extract trace data.
Our validation approach features a designed simulation experiment in which trace data associated with a changing subset of input processes are fixed---which we refer to as a \textit{subtrace}, while inputs from the remaining processes are simulated using their estimated counterparts. This technique enables us to obtain output distributions conditional not only on the state of the system, but also on subtraces. Existing validation approaches either fix the trace data from all input processes \cite{kleijnen1998validation} or simulate using only estimated input models \cite{rhodes2023tracking}. In light of the output distributions on which these methods validate a simulation model, we will refer to our approach as \textit{conditional} validation and other existing approaches as \textit{marginal} validation.
Relative to marginal validation, conditional validation tests a more comprehensive notion of alignment and therefore provides a modeler with stronger assurances about the validity of their simulation model.

As we will show, conditional validation provides a principled way for validating a simulation model conditional on different components of the system and assessing how different input models contribute to model misalignment.
Our approach utilizes the PIT technique of \cite{rhodes2023tracking}---this time applied to conditional distributions conditioned on subtraces---and performs multiple hypothesis testing. For diagnosing specific sources of misalignment, we employ bootstrapping in full and fractional factorial designs to systematically explore the design space \cite{wu2011experiments}. Stepwise variable selection and regression tree-based importance measures are then applied to the generated dataset to attribute observed deviations to specific input models.

The remainder of the paper is organized as follows. In Section~\ref{sec:cond_val}, we formalize our validation framework and define subtrace-conditional output distributions and corresponding hypothesis tests. In Section~\ref{sec:sources}, we present a design-of-experiments approach to analyze the effects of different input models and diagnostic methods that attribute model misalignment to individual sources of uncertainty. We present numerical experiments in Section~\ref{sec:exp} and conclude in Section~\ref{sec:conc}.

\section{SUBTRACE-CONDITIONAL VALIDATION}
\label{sec:cond_val}
We consider a stochastic system that evolves dynamically over discrete time periods $t = 1, 2, \dots, T$. At each period $t$, let $\mathbf{\psi}_t$ denote the quantities known, observed, or decided at the beginning of period $t$ that influence the evolution of the system over period $t$. For example, $\psi_t$ could consist of state variables and actions taken at the beginning of period $t$. Let $\Xi_t = (\Xi_{t1}, \Xi_{t2}, \dots, \Xi_{tS})$ denote the collection of random primitives that drive the behavior of the system over the subsequent period, where $S$ is the number of distinct input processes. These quantities reflect \emph{aleatoric uncertainty}, i.e., uncertainty associated with events that will occur in period $t$ but are not yet realized at the beginning of period $t$. We consider a scalar system output given by $Y_t = H(\mathbf{\psi}_t, \Xi_t)$, where $H$ captures the dynamics of the system.
We assume that the simulation model being studied correctly specifies the mapping $H$, so that any misalignment between the simulated outputs and the real-world outputs arises solely from misspecification of the probability model for $\Xi_t$. This assumption is commonly made in other areas of the simulation literature, such as input uncertainty, where system dynamics are assumed to be modeled correctly and the focus is on quantifying the impact of stochastic input models on output variability \cite{song2014advanced}.

As a concrete example, in a call center simulation, 
the observable state $\mathbf{\psi}_t$ might consist of the number of customers in the queue and the number of agents available at the beginning of period $t$;
the random primitives $\Xi_t$ might consist of the interarrival times and 
service times during period $t$, corresponding to distinct input processes; and the system output $Y_t$ may represent some performance metric of interest, such as the average 
waiting time or the number of callers who wait longer than one minute during period $t$.

We assume access to a historical trajectory $\mathcal{E} = \{ (\mathbf{\psi}_t, \xi_t, y_t) \}_{t=1}^T$, where $y_t$ denotes the observed output in period $t$, and $\xi_t = (\xi_{t1}, \xi_{t2}, \dots, \xi_{tS})$ denotes the corresponding realization of $\Xi_t$, i.e., the trace for period $t$.
More specifically, $\xi_{ts}$ represents a vector of the inputs associated with the $s$th input process, $s = 1, 2, \dots, S$, during period $t$, i.e., the realization of $\Xi_{ts}$. The length of $\xi_{ts}$ may vary from period to period.

Let $\mathcal{A}$ be the set of all subsets of $\{1,2,\ldots, S\}$.
For any period $t$ and subset $A \in \mathcal{A}$, define $\Xi_{t,A} = \{\Xi_{ts} \colon s \in A\}$ and let $\xi_{t,A}$ denote its realized value, which we refer to as the \textit{subtrace} associated with $t$ and $A$.
We consider a family of conditional output distributions characterized by fixing a subtrace and sampling from the other input processes. 
More precisely, for any fixed $t$ and $A$, we define the \textit{subtrace-conditional output distribution}
\[
F_{t,A}(y) := \mathbb{P}\left( Y_t \le y \mid \mathbf{\psi}_t, \Xi_{t,A} = \xi_{t,A} \right) \text{ for all } y \in \mathbb{R}.
\]
This distribution represents the law of the system output when the components of randomness indexed by $A$ are fixed to their historical values, and the remaining components $\Xi_{t,A^c}$ are random. 

\textbf{Remark:} There is some subtlety to the previous statement, as it presumes that the primitives that comprise the subtrace $\xi_{t, A}$ are the only ones from the input processes in $A$ needed to run a replication of the simulation model, regardless of the realization of $\Xi_{t, A^c}$. However, for many simulation models, this will not be the case. For example, in a simple queueing system, if we were to reuse historical service times, but then generated more arrivals than occurred that day, we would need to generate additional service times within the simulation. Our framework can still be applied in such settings, but greater care is needed when interpreting the subtrace-conditional output distributions and when designing the simulation code to accommodate such situations.

\subsection{Probability Integral Transformation}

The challenge in validating a simulation model in this digital-twin setting is twofold: First, we observe only a single realization of the triplet $(\psi_t, \xi_t, y_t)$ for each time period $t$. Second, the conditional distribution of $Y_t$---conditioned on $\psi_t$ and, in our case, parts of $\Xi_t$---is different for different $t$.
To address this, we adopt the PIT technique of \cite{rhodes2023tracking}. 
For each $t = 1, 2, \dots, T$ and each $A \in \mathcal{A}$, we define a variable $U_{t,A} := F_{t,A}(y_t)$.
Assuming that the output is continuous-valued and that the time periods are sufficiently long so that the outputs across different periods are conditionally independent, conditioned on $\{\psi_t\}_{t=1}^T$, it follows that
\[
\{U_{t,A}\}_{t=1}^T \overset{\mathrm{i.i.d.}}{\sim} U(0,1),
\]
where $U(0, 1)$ denotes the continuous uniform distribution on $[0, 1]$.
This application of the PIT enables us to pool the transformed variables $\{U_{t,A}\}_{t=1}^T$ to validate the simulation model with respect to each subset $A$. 

In practice, the true input processes are unknown and are modeled using historical data, meaning that the subtrace-conditional output distribution $F_{t,A}$ is generally unavailable.
However, $F_{t,A}$ can be approximated via simulation by sampling $R$ i.i.d.\ realizations of ${\Xi}_{t,A^c}$ from the corresponding fitted input models, denoted by $\Xi_{t,A^c}^{1}, \Xi_{t,A^c}^{2}, \dots, \Xi_{t,A^c}^{R}$, and using them---alongside the subtrace $\xi_{t,A}$---to drive the simulation model and obtain i.i.d.\ outputs $Y_{t,A}^{1}, Y_{t,A}^{2}, \ldots, Y_{t,A}^{R}$, where $Y_{t,A}^{r} = H(\mathbf{\psi}_t,\xi_{t,A}, \Xi_{t,A^c}^{r})$ for $r = 1, 2, \ldots, R$.
(Although the notation $H(\mathbf{\psi}_t,\xi_{t,A}, \Xi_{t,A^c}^{r})$ gives the impression that the only primitives from input processes in $A$ needed to execute a simulation replication are those contained in the subtrace $\xi_{t,A}$, as previously discussed in the remark above, the framework extends to settings in which it is necessary to generate additional primitives from the associated input models.)
The corresponding empirical cumulative distribution function (ecdf)
\[
\hat{F}_{t,A}(y) = \frac{1}{R} \sum_{r=1}^R \mathds{1}\{Y_{t, A}^{r} \leq y\} \text{ for all } y \in \mathbb{R}
\] 
can then be used to estimate $U_{t, A}$ by $\hat{U}_{t,A} := \hat{F}_{t,A}(y_t)$.

\subsection{Hypothesis Testing}

For a fixed $A \in \mathcal{A}$, we consider the hypothesis test
\[
H_{0, A}\colon \{U_{t,A}\}_{t=1}^T \overset{\text{i.i.d.}}{\sim} U(0,1)
\quad \text{versus} \quad
H_{1,A}\colon \{U_{t,A}\}_{t=1}^T \overset{\text{i.i.d.}}{\not\sim} U(0,1).
\]
As a test statistic, we measure the discrepancy between the empirical distribution of $\{\hat{U}_{t,A}\}_{t=1}^T$, denoted by $\tilde{F}_{A}$, and the distribution $U(0, 1)$ using the Wasserstein-1 distance,
\[
W_A := W_1\big(\tilde{F}_{A}, U(0,1)\big)
= \inf_{\pi \in \Pi(\tilde{F}_{A}, U(0,1))} 
\int_{[0, 1]^2} |u - v| \, d\pi(u,v),
\]
where $\Pi(\tilde{F}_{A}, U(0,1))$ denotes the set of all joint probability measures (couplings) on $[0,1]^2$ with marginals $\tilde{F}_{A}$ and $U(0,1)$.
The Wasserstein-1 distance captures the average deviation across the distribution and is less sensitive to local fluctuations compared to measures such as the Kolmogorov-Smirnov test statistic \cite{villani2009optimal}.
In our setting, $W_A$ is the optimal value of a semi-discrete optimal transport problem, which, in one dimension, admits a closed-form solution \cite{vallender1974}, namely,
\[
W_A
= \frac{1}{T} \sum_{t=1}^T
\left| \hat{U}_{(t), A} - \frac{2t-1}{2T} \right|,
\]
where $\hat{U}_{(1), A} \le \hat{U}_{(2), A}\le \cdots \le \hat{U}_{(T), A}$ denote the order statistics of $\{\hat{U}_{t,A}\}_{t=1}^T$.
Larger values of $W_A$ indicate a greater deviation from uniformity and provide evidence that the input models in $A^c$ may be misspecified. (For ease of notation and exposition, we will refer to carrying out the simulation experiment above \emph{for all $A \in \mathcal{A}$} or for all $A$ in some design set, but note that for the case $A = \{1,2,\ldots, S\}$, $W_A$ is trivially zero, since the mapping $H$ is assumed to be correctly specified.)

A $p$-value for this hypothesis test can be obtained via Monte Carlo or bootstrapping by comparing the observed test statistic to its distribution under $H_{0, A}$. The null hypothesis is rejected at a significance level $\alpha$ if $p_{A} \le \alpha$, where $p_{A}$ is the $p$-value for the test.

Because validation is to be performed across all $A \in \mathcal{A}$, a multiple hypothesis testing procedure can be applied where the Bonferroni or Holm correction is used to control the family-wise error rate. With the Bonferroni correction, for example, we would reject $H_{0, A}$ if $p_{A} \le \alpha/(|\mathcal{A}|-1)$.
This test is related to the marginal validation procedure introduced in \cite{rhodes2023tracking}. Specifically, this approach consists of $2^{|S|}-1$ hypothesis tests, excluding $A = \{1,2,\ldots, S\}$.
Marginal validation arises as a special case within our broader conditional validation framework, corresponding to testing with $A = \emptyset$, in which case all input processes are simulated. 
As a result, passing our validation procedure provides a stronger certificate that the model is valid than passing the test in \cite{rhodes2023tracking}.
Moreover, the pattern of rejected and non-rejected hypotheses across $\mathcal{A}$ provides valuable diagnostic information to the modeler. For example, rejection of one of the tests in which a particular input process is simulated suggests that the corresponding input model may be misspecified. In Section~\ref{sec:sources}, we discuss procedures that can help the modeler identify and quantify the contributions of individual input processes to overall model misalignment.
We also note that our approach is amenable to parallelization, both for calculating $W_A$ across different subsets $A \in \mathcal{A}$, as well as executing simulation replications at a given $A$.

\subsection{Streaming Data}

In many practical settings, data are not available in a static batch, but rather arrive sequentially over time. The validation and diagnostic analysis described in this paper can be updated or re-run as new observations from the physical system become available.
Suppose that at the end of period $t$ we have observed data 
\[
\mathcal{E}_t = \{(\mathbf{\psi}_\tau, \xi_\tau, y_\tau)\}_{\tau=1}^t,
\]
and have calculated the discrepancy measures $\{W_{A}\}_{A \in \mathcal{A}}$. At the end of period $t+1$, when the triplet $(\mathbf{\psi}_{t+1}, \xi_{t+1}, y_{t+1})$ has been fully observed, we can augment the previous dataset to form
\[
\mathcal{E}_{t+1} = \mathcal{E}_t \cup \{(\mathbf{\psi}_{t+1}, \xi_{t+1}, y_{t+1})\}.
\]
With modest additional simulation effort, we can obtain $\hat{U}_{t+1, A} = \hat{F}_{t+1,A} (y_{t+1})$ for each $A \in \mathcal{A}$ and add these values to their respective sets and recompute the discrepancy measures $\{W_{A}\}_{A \in \mathcal{A}}$.

This updating procedure requires running additional simulation replications for only the latest period. %newly observed state and output.
Relative to the computational cost of these replications, the ensuing update of the discrepancy measure is inexpensive. As a result, the framework could be adapted to employ sequential probability ratio tests \cite{wald1947} and thereby enable a dynamic validation pipeline in which one can track the validity of the simulation model and how the contributions of different input models to misalignment evolve over time.

\subsection{Classical Simulation Setting} 
The proposed framework can also be applied to validate simulation models whose initial conditions and input models do not depend on time.
With some abuse of notation, in this setting, we assume that we have access to a historical trajectory $\{\xi_t, y_t\}_{t=1}^T$, where $y_t = H(\xi_t)$, and $t$ now denotes the observation index.
Furthermore, we assume that the random primitives $\Xi_1, \Xi_2, \ldots, \Xi_T$ are mutually independent, hence, so are the outputs $Y_1, Y_2, \ldots, Y_T$.
For any subset $A \in \mathcal{A}$, we again consider conditional distributions of $Y$ conditional on fixing the subtrace $\xi_{t,A}$ and simulating the remaining inputs $\Xi_{t, A^c}$. This yields a collection of subtrace-conditional output distributions \textit{across observations}, to which the validation approach outlined above can be straightforwardly applied.
This setting differs slightly from the digital-twin setting in terms of the computational cost. In particular, for $A = \emptyset$, i.e., when all inputs are simulated, we need only run $R$ replications of the simulation model to estimate the distribution of $Y$, because this distribution no longer depends on a time-varying system state. The methods introduced in Section~\ref{sec:sources} can likewise be adapted to this setting.

\section{DETECTING THE SOURCE(S) OF MISALIGNMENT}
\label{sec:sources}
The multiple hypothesis testing framework can help determine \emph{whether} the simulation model is statistically valid. The goal of this section is instead to identify \emph{which input models} are responsible for any observed misalignment. To address this question, we leverage ideas from the design of experiments (DoE) and bootstrapping to construct a dataset that can then be analyzed using statistical learning techniques.

\subsection{Design of Experiment}
\label{sec:doe}

A natural question is how to systematically obtain and analyze the data generated by simulation experiments like those described in Section~\ref{sec:cond_val} to understand \emph{which} sources of uncertainty contribute the most to model misalignment.
This motivates the use of DoE. We consider the dataset generated from a designed experiment on a design set $\mathcal{D} \subseteq \mathcal{A}$, $\left\{ \big(\mathbf{z}_{A}, W_{A}\big) \colon A \in \mathcal{D} \right\}$, where $\mathbf{z}_{A} \in \{0,1\}^S$ is a binary encoding of subset $A$, with
\[
z_{s,A} =
\begin{cases}
0 & \text{if inputs from input process $s$ are simulated},\\
1 & \text{if inputs from input process $s$ are fixed to their observed values}.

\end{cases}
\]
We refer to a design point associated with a vector $\mathbf{z}_{A}$ as a \textit{configuration}.

We first consider a full factorial design in which all $2^S$ subsets of $\{1, 2, \dots, S\}$ are evaluated, i.e., $\mathcal{D} = \mathcal{A}$.
The full factorial design supports the use of analysis of variance (ANOVA) \textit{without replication} to quantify the contributions of individual factors and a subset of their interactions \cite{oehlert2010first}. Graphical tools such as response grid plots (RGPs) \cite{barton2025response} can also provide intuitive visualizations of these effects.
A significant main effect suggests that the corresponding input model is a primary driver of misalignment. Interaction effects, on the other hand, capture whether the impact of one input model on overall misalignment depends on the status of another input model. A significant interaction effect indicates that deviations arise from the joint misspecification of multiple input models.

In situations where evaluating all $2^S $ subsets is computationally infeasible, we can instead employ fractional factorial designs to efficiently explore the design space while substantially reducing the number of configurations \cite{wu2011experiments}.
Fractional factorial designs feature a structured subset of design points that enables the estimation of all main effects and select low-order interactions. These designs introduce aliasing between effects; however, this issue can be managed using standard design principles. In particular, one can invoke the \emph{hierarchical effect principle}, which assumes that higher-order interactions are unlikely to be significant unless their corresponding lower-order effects are also significant.
Under this principle, one can focus on estimating the main effects and low-order interactions, which are typically the most interpretable and practically relevant. If needed, additional runs can be performed to resolve important aliases and refine the analysis.

To quantify the uncertainty in our estimates of the Wasserstein distance between the distribution of $\{U_{t,A}\}_{t=1}^T$ and $U(0, 1)$ due to the choice of $R$, we propose leveraging bootstrapping to enrich the dataset $\left\{ \big(\mathbf{z}_{A}, W_{A}\big) \colon A \in \mathcal{D} \right\}$. 
For each $t = 1, 2, \ldots, T$ and $A \in \mathcal{D}$,  we first obtain $B$ bootstrapped datasets by sampling with replacement from $\{\hat{Y}_{t,A}^{r}\}_{r=1}^R$.
Let $\{\hat{Y}_{t,A}^{r,b}\}_{r=1}^R$ denote the $b$th bootstrapped dataset, for $b = 1, 2,\dots, B$. For each $b$, we then construct the corresponding empirical distribution $\hat{F}_{t,A}^{b}$ and compute $\hat{U}_{t, A}^{b} := \hat{F}_{t, A}^{b}(y_t)$.
By pooling the values $\{\hat{U}_{1, A}^{b}, \hat{U}_{2, A}^{b}, \ldots, \hat{U}_{T, A}^{b}\}$, we obtain a bootstrapped realization of the test statistic $W_{A}^{b}$ for $b = 1,2,\dots,B$.
Repeating the process for all $A \in \mathcal{D}$, we produce an enlarged dataset $\{(\mathbf{z}_A, W_A^{b})\colon A \in \mathcal{D},\, b \in \{1,2,\ldots, B\}\}$ which we can then use for our analysis.
Importantly, generating this aggregated bootstrapped dataset does not require running more simulation replications.

\subsection{Subset Selection}

One approach to quantifying the contribution of individual input models to overall misalignment is to frame a regression problem where the Wasserstein-1 discrepancy measure is the response and the statuses of the input models are the predictors. Specifically, we leverage regression for \textit{inference} rather than \textit{prediction}, i.e., to see which input models explain the variation in $W$. Consider the model
\[
W_A = \beta_0 + \sum_{s=1}^S \beta_s z_{s,A} + \sum_{s < s'} \beta_{ss'} z_{s,A} z_{s',A} + \varepsilon_{A},
\]
where the terms $\beta_s$ represent the main effects, $\beta_{ss'}$ capture the interaction effects between pairs of input processes and $\varepsilon_A$ is the mean-zero noise. When using the original dataset $\{(\mathbf{z}_A, W_A)\colon A \in \mathcal{D}\}$, $\varepsilon_A$ is equal to the net effect of effects of order higher than two. When using the dataset resulting from bootstrapping $\{(\mathbf{z}_A, W_A^{b})\colon A \in \mathcal{D},\, b \in \{1,2,\ldots, B\}\}$, it also captures the uncertainty originating from having run a finite number of simulation replications.

For cases where $S$ is large, we propose employing \textit{subset selection} procedures such as forward selection and backward stepwise selection to identify a subset of predictors (input models) that explains the variability in $W_{A}$.

Forward stepwise selection begins with an intercept-only model (i.e., the null model) and sequentially adds input models or their interactions to produce a sequence of nested models. At each step, the procedure adds the input model or interaction that yields the greatest improvement in a chosen criterion (e.g., Akaike Information Criterion (AIC), Bayesian Information Criterion (BIC), or adjusted $R^2$) among all candidate input models or interactions not yet included in the model (including interaction terms, if considered). The procedure continues until no additional input model or interaction provides a statistically or practically meaningful improvement.
Input models or interactions selected early in the procedure can be interpreted as dominant sources of model misalignment.

Backward stepwise selection starts from a model that includes all input models and their interactions and iteratively removes them. At each step, the input model or interaction whose removal results in the smallest deterioration in model fit (or the largest improvement according to a penalized criterion such as AIC or BIC) is eliminated. The process continues until all remaining input models or interactions are statistically significant or contribute meaningfully to explaining the variability in $W_{A}$.
This approach is most appropriate when we suspect that all or most of the input models are misspecified and the goal is to remove redundant or non-influential ones. Input models or interactions removed early in the elimination process can be interpreted as having a negligible impact on model misalignment.

\subsection{Regression Trees}

We can alternatively model the relationship between the configuration $\mathbf{z}_{A}$ and $W_{A}$ using a regression tree.
Here, too, our interest in regression trees is for making inference rather than for making predictions.
A regression tree is trained to predict $W_{A}$ from $\mathbf{z}_{A}$ using recursive binary splitting. At each split, the algorithm selects the input model to cut on---separating based on its status (historical vs simulated)---that maximally reduces the residual sum of squares (RSS). When using the original dataset, the total variability in the response is measured by the total sum of squares (TSS),
$$
\mathrm{TSS} = \sum_{A \in \mathcal{D}} \left( W_{A} - \bar{W} \right)^2, \text{ where }
\bar{W} = \frac{1}{|\mathcal{D}|} \sum_{A \in \mathcal{D}} W_{A},
$$
and the final RSS in this case will be zero.
When using the bootstrapped dataset we have 
$$
\mathrm{TSS} = \sum_{A \in \mathcal{D}} \sum_{b=1}^{B} \left( W^b_{A} - \bar{\bar{W}} \right)^2, \text{ where }
\bar{\bar{W}} = \frac{1}{|\mathcal{D}|B} \sum_{A \in \mathcal{D}}\sum_{b=1}^{B} W^b_{A},
$$
and the final RSS is
$$
\mathrm{RSS} = \sum_{A \in \mathcal{D}} \sum_{b=1}^{B} (W^b_{A} - \bar{W}_{A})^2, \text{ where } \bar{W}_{A} = \frac{1}{B}\sum_{b=1}^B W^b_{A}.
$$

A variable importance score for each source $s$, $s = 1, 2, \dots, S$, can be calculated in terms of the total reduction in RSS attributable to splits on that source, aggregated over the entire tree. These importance scores essentially decompose the variability in $W_{A}$ across input models, with higher importance scores indicating that the associated input models are likely more responsible for model misalignment. This approach resembles that of \cite{ghasemloo2025quantifying} for decomposing (epistemic) input uncertainty across input models.
The approach also parallels ANOVA, as it decomposes the variability of the response into contributions from different factors. However, unlike classical ANOVA, which requires explicit specification of interaction terms, regression trees capture interactions implicitly through hierarchical splitting. This is particularly advantageous in applications where one is interested in apportioning the contributions to individual predictors.

\section{Numerical Experiments}
\label{sec:exp}
We illustrate the proposed framework through two numerical experiments.
In the first experiment, we study a classical simulation model of an $M/M/1$ queue to highlight the importance of validating at the level of subtrace-conditional output distributions. We demonstrate that relying solely on marginal validation metrics can fail to detect input model misspecification when opposing biases from misspecified input models offset each other. We show that these lower-level deviations can be effectively revealed by our conditional approach to validation.
In the second experiment, we apply the proposed framework to a digital-twin-enabled simulation model of a tandem queueing system. 

\subsection{$M/M/1$ Queue}

We consider an $M/M/1$ queueing system in which the KPI of interest is customers' mean waiting time in the queue. It is well known that the expected value of the steady-state mean waiting time depends on the difference $\mu - \lambda$, where $\lambda$ and $\mu$ denote the arrival and service rates, respectively \cite{shortle2018fundamentals}. This relationship creates a setting in which different parameter pairs of the arrival process and service time distribution can yield similar output behavior. 
In our experiment, we study a terminating simulation of an $M/M/1$ queue over a finite horizon of 600 minutes with the true underlying parameters set to $\lambda = 1  \text{~min}^{-1}$ and $\mu = 2 \text{~min}^{-1}$. Using these parameters, we perform $40$ independent simulation replications, starting from an empty queue, and recording the corresponding mean waiting times.
To illustrate the potential for marginal validation approaches to incorrectly validate a misspecified model, we consider such a model instance with parameters $\hat{\lambda} =1.5  \text{~min}^{-1}$ and $\hat{\mu} = 2.5  \text{~min}^{-1}$. Both the ``true'' system and the misspecified model satisfy $\mu - \lambda = \hat{\mu} - \hat{\lambda} = 1 \text{~min}^{-1}$.

We consider three design points of (arrival process, service times): $(0,0)$, $(0,1)$, and $(1,0)$, where 0 indicates that that input process is simulated and 1 indicates that that input process is fixed to historical values. (The design point $(1, 1)$ is not simulated because it corresponds to reusing all historical inputs.) At each design point, we run $R=40$ simulation replications, construct the ecdf of the mean waiting time, and apply the PIT to the 40 historical observations of the mean waiting time.
These values are then pooled for $t = 1, 2, \ldots, 40$ and used to compute the discrepancy measure $W_A$ along with the corresponding $p$-value. The results are summarized in Table~\ref{tab:mm1_validation}.

\begin{table}[ht]
\centering
\caption{Wasserstein-1 distances ($W_A$) and $p$-values ($p_A$) for design points simulated to validate a simulation model of an M/M/1 queue.}
\label{tab:mm1_validation}
\begin{tabular}{cc|cc}
%\hline
Arrival Process & Service Times & $W_A$ & $p_A$ \\
\hline
Simulated & Simulated & 0.038 & $0.758$ \\
Historical & Simulated & 0.47 & $< 0.001$ \\
Simulated & Historical & 0.5 & $< 0.001$
%\hline
\end{tabular}
\end{table}

We observe that when both the arrival and service processes are simulated, the Wasserstein-1 distance is small, suggesting that the model appears reasonably well aligned under the marginal validation approach. However, the misalignment becomes evident when we reuse subtraces for the arrival or service-time processes. In particular, reusing the historical arrival times or the historical service times leads to substantially larger Wasserstein-1 distances and extremely small $p$-values, providing strong evidence that both input models are misspecified.

These results highlight a limitation of marginal validation: opposing biases induced by misspecified input models---in this case, in the parameter settings of $\lambda$ and $\mu$---can offset each other in such a way that the marginal output distribution appears consistent with the observed output data. In contrast, the conditional approach to validation can correctly detect that multiple input models are misspecified.
This mode of failure for marginal validation methods is especially critical in practical settings where the (invalid) simulation model will subsequently be used to evaluate different system configurations, i.e., different mappings $H$, but with the same (misspecified) input models, e.g., when performing simulation optimization.

\subsection{Tandem Queueing System}

In the second experiment, we apply the proposed framework to a tandem queueing system. This setting is designed to mimic a multi-stage service process with congestion, blocking, and time-varying arrivals, while also supporting hot-start initialization from observed system states.
We consider a tandem queueing system consisting of 3 sequential service stations, where jobs arrive at the first station and must be processed sequentially at all three stations before exiting the system. Each station has a single dedicated server (station) and an associated finite buffer. The buffer capacities are given by $(10, 15, 15)$.
The system is subject to blocking: if a job completes service at Station $i$, $i = 1, 2$, but the buffer at Station $i+1$ is full, the server at Station $i$ becomes blocked and cannot release the job until space becomes available. This induces upstream congestion and complex system dynamics. A job balks, i.e., does not enter the system, if the first station's buffer is full.
Service times at each station are assumed to be independent and exponentially distributed with mean service times of 1, 1, and 2 minutes, respectively.
Arrivals follow a non-stationary Poisson process with piecewise-constant rates over 30-minute intervals, as provided in Table~\ref{tab:arrival_rates}.
\begin{table}[ht]
\centering
\caption{Arrival rates per 30-minute interval for a 12-hour day.}
\label{tab:arrival_rates}

\scriptsize
\setlength{\tabcolsep}{1.5pt}

\begin{tabular}{c|cccccccccccccccccccccccc}
%\hline
Interval & 1 & 2 & 3 & 4 & 5 & 6 & 7 & 8 & 9 & 10 & 11 & 12 & 13 & 14 & 15 & 16 & 17 & 18 & 19 & 20 & 21 & 22 & 23 & 24 \\
\hline
Rate (min$^{-1}$)& 1.18 & 1.23 & 1.21 & 1.17 & 1.24 & 1.19 & 1.22 & 1.16 & 1.20 & 0.98 & 1.03 & 1.01 & 0.97 & 1.04 & 0.99 & 1.02 & 0.96 & 1.00 & 1.05 & 1.08 & 1.02 & 0.99 & 1.01 & 1.06\\
%\hline
\end{tabular}

\normalsize
\end{table}

The system is simulated over a horizon of 720 minutes (12 hours) partitioned into 24 intervals of 30 minutes each. The KPI of interest is the fraction of balked jobs and is recorded at the end of each interval throughout a single 12-hour replication. The system state at any time $t$ is fully characterized by the number of jobs in each buffer, the status of each station (idle, processing, or blocked), and the elapsed service time of the jobs currently in service.

In our validation experiments, we assume that the non-stationary arrival process and the service time distributions for Stations 1 and 3 have been exactly estimated, but the mean service time for Station 2 has been erroneously estimated as 2 (compared to its true value of 1).
We execute the validation procedure over all $2^4 = 16$ possible configurations. For each configuration other than the configuration that reuses the full trace, we perform $R = 50$ independent simulation replications for each of the $T = 24$ intervals, each time using the historical state of the system, $\psi_t$. We compute the ecdfs $\hat{F}_{t, A}$ for each $t$ and $A$ and the values $\{\hat{U}_{t, A}\}_{t=1}^{T=24}$ for each $A$.
We perform bootstrapping with $B=20$ on the original dataset to obtain the bootstrapped dataset described at the end of Section~\ref{sec:doe}.

\begin{figure}[tbh]
    \centering
    \begin{subfigure}[t]{0.48\textwidth}
        \centering
        \includegraphics[width=\textwidth]{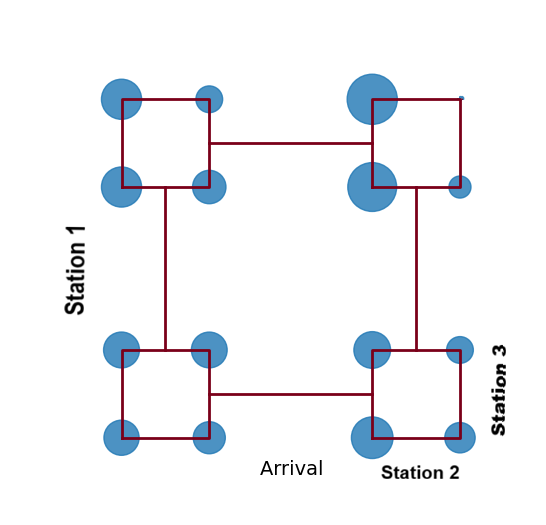}
        \caption{Wasserstein-1 distance.}
        \label{fig:rgp_ks}
    \end{subfigure}
    \hfill
    \begin{subfigure}[t]{0.44\textwidth}
        \centering
        \includegraphics[width=\textwidth]{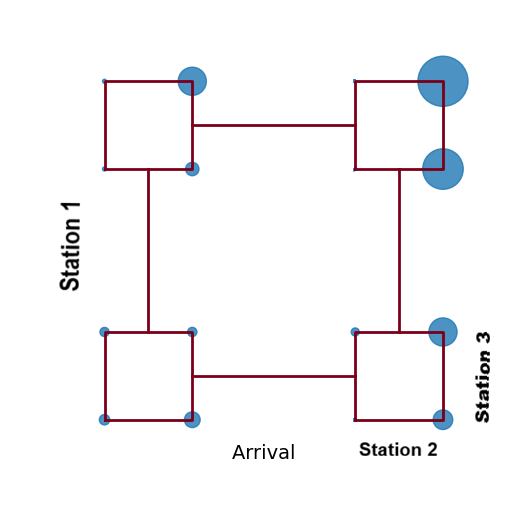}
        \caption{$p$-value.}
        \label{fig:regp_pval}
    \end{subfigure}

    \caption{Response grid plots (RGP) for a full factorial with $2^4$ design of the input models (simulated (low) vs historical (high)). The radius of each circle indicates the relative magnitude.}
    \label{fig:rgp}
\end{figure}

\begin{figure}[!htbp]
    \centering

        \includegraphics[width=0.8\textwidth]{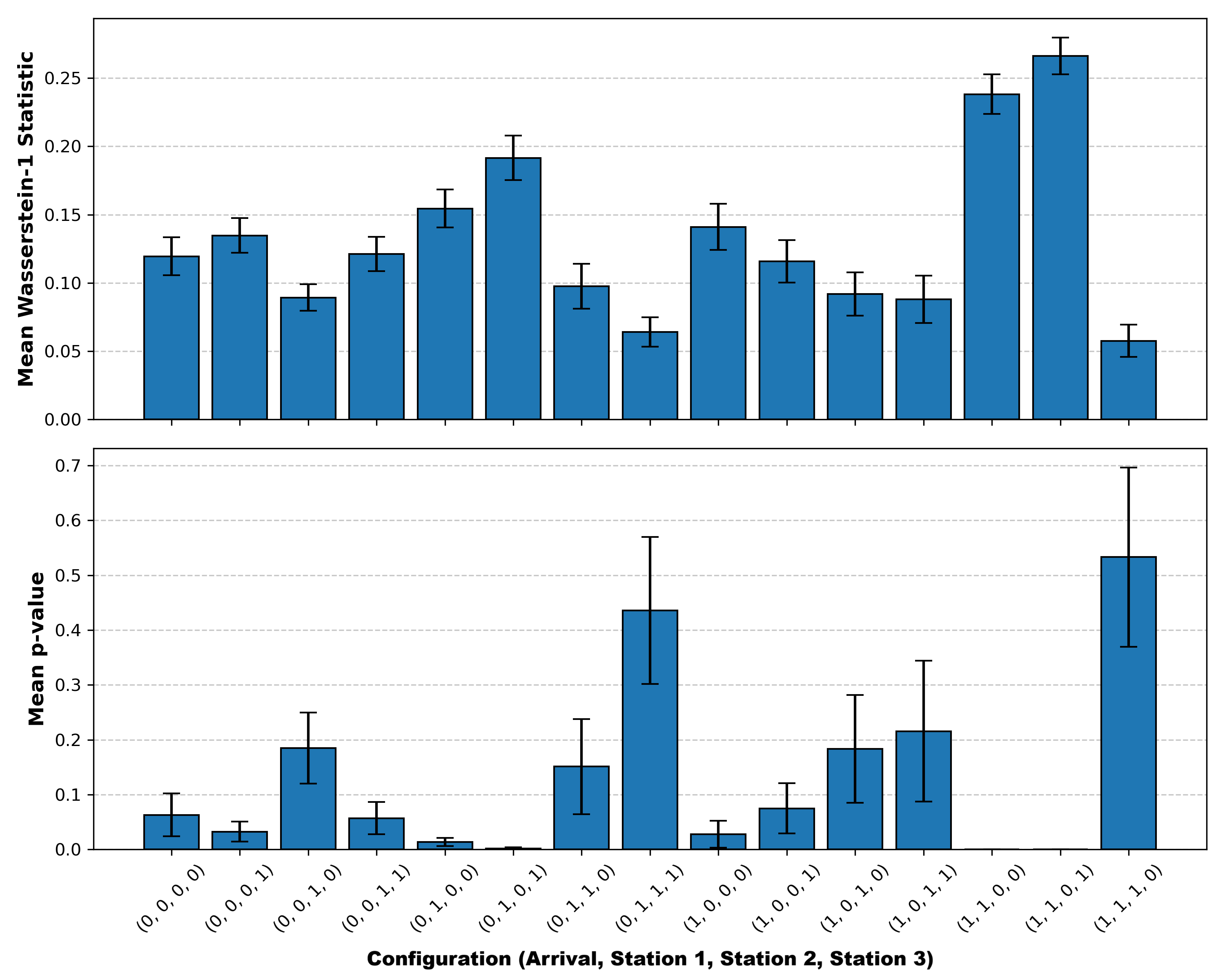}
        \caption{Wasserstein-1 distance and corresponding $p$-values for different configurations in trace-conditional validation analysis for a misspecified tandem queueing system.}
        \label{fig:config_ks}
   
\end{figure}

\begin{figure}[!htbp]

        \centering
        \includegraphics[width=0.8\textwidth]{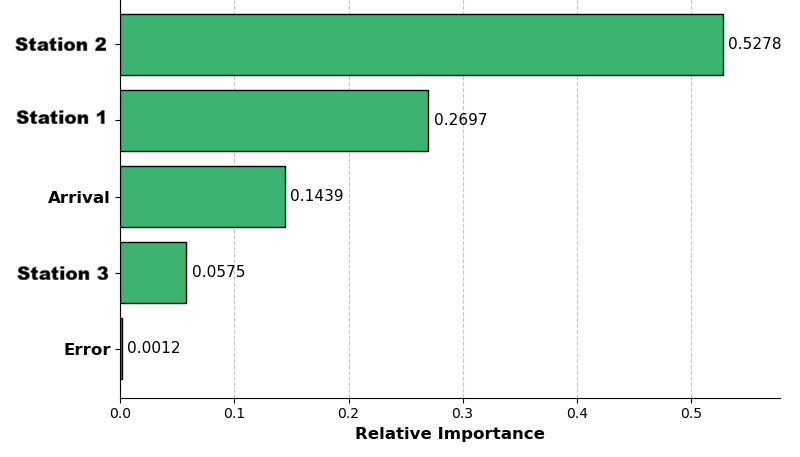}
        \caption{Variable importance scores from decomposing the variation in the Wasserstein-1 discrepancy measure across different input models by trace-conditional validation analysis for a misspecified tandem queueing system.}
        \label{fig:decomp}

\end{figure}

Response grid plots of the Wasserstein-1 distance and the associated $p$-values for each configuration are depicted in Figure \ref{fig:rgp} and corresponding mean values in Figure~\ref{fig:config_ks}. We observe that configurations in which the subtrace contains historical service times from Station 2 exhibit significantly lower discrepancies, since the primary source of misspecification is absent. Figure \ref{fig:decomp} shows the importance scores obtained by training a regression tree, where the statuses of the input models are used as predictors and the Wasserstein-1 distance is the response. The results indicate that the service time distribution of Station 2 is the primary contributor to model misalignment. Moreover, the magnitude of the residual (error) term suggests that the number of simulation replications used in the analysis, $R$, is sufficient to reliably detect the sources of misalignment.


\newpage
\section{CONCLUSION}
\label{sec:conc}
We propose a statistical validation framework for simulation models and digital twins that validates not only the marginal distribution of the output, but also a family of related conditional distributions that condition on partial historical trace data. We provide interpretable tools for identifying sources of model misspecification and their interactions through the integration of design of experiments, subset selection, and regression trees. Numerical experiments demonstrate that the proposed approach can uncover hidden misalignments that are not detectable through marginal validation methods and quantify the contribution of each input model to the overall misalignment. Future research could explore multi-dimensional outputs and more complex situations such as censored trace data.
\section*{ACKNOWLEDGMENTS}
\label{sec:ack}

This work was supported by the Intuit University Collaboration Program and National Science Foundation Grant OAC-2410948. We also thank Dusan Bosnjakovic and David Afshartous for their support throughout the project and Barry Nelson for helpful conversations.

% Reducing font size (to 9pt) for References & Author Biagraphies
\footnotesize

% Please don't exchange the bibliographystyle style
\bibliographystyle{plain}

% AUTHOR: Include your bib file here
\bibliography{mainbib}

\section*{AUTHOR BIOGRAPHIES}

\noindent {\bf \MakeUppercase{Mohammadmahdi Ghasemloo}} is a PhD student in the Wm Michael Barnes '64 Department of Industrial and Systems Engineering at Texas A\&M University.
His research interests lie at the intersection of machine learning and stochastic simulation optimization. His e-mail address is \email{mohammad\_ghasemloo@tamu.edu}.\\

\noindent {\bf \MakeUppercase{David J. Eckman}} is an Assistant Professor in the Wm Michael Barnes '64 Department of Industrial and Systems Engineering at Texas A\&M University.
His research interests deal with optimization and output analysis for stochastic simulation models.
He is a co-creator of SimOpt, a testbed of simulation optimization problems and solvers.
His e-mail address is \email{eckman@tamu.edu}.\\

\noindent {\bf \MakeUppercase{Yaxian Li}} is a Staff AI Scientist at Intuit within the A2D department. Her research interests encompass simulation, optimization, artificial intelligence and machine learning. Her e-mail address is \email{Yaxian\_Li@intuit.com}.\\ 

\end{document}